# Dielectric Properties of Metal-Organic Frameworks Probed via Synchrotron Infrared Reflectivity

Matthew R. Ryder,[1,2,3] Zhixin Zeng,[1] Yueting Sun,[1] Irina Flyagina,[1] Kirill Titov,[1] E.M. Mahdi,[1]
Thomas D. Bennett,[4] Bartolomeo Civalleri,[5] Chris S. Kelley,[2] Mark D. Frogley,[2]
Gianfelice Cinque,[2] and Jin-Chong Tan[1*]

[1]Multifunctional Materials & Composites (MMC) Laboratory, Department of Engineering Science, University of Oxford, Parks Road, Oxford OX1 3PJ, United Kingdom
[2]Diamond Light Source, Harwell Campus, Oxford OX11 0DE, United Kingdom
[3]ISIS Facility, Rutherford Appleton Laboratory, Chilton, Didcot OX11 0QX, United Kingdom
[4]Department of Materials Science and Metallurgy, University of Cambridge, Cambridge CB3 0FS, United Kingdom
[5]Department of Chemistry, NIS and INSTM Reference Centre, University of Turin, via Pietro Giuria 7, 10125 Torino, Italy

**Abstract:** We present the frequency-dependant (dynamic) dielectric response of a group of topical polycrystalline zeolitic imidazolate-based metal-organic framework (MOF) materials in the extended infrared spectral region. Using synchrotron-based FTIR spectroscopy in specular reflectance, in conjunction with density functional theory (DFT) calculations, we have revealed detailed structure-property trends linking the THz region dielectric response to framework porosity and structural density. The work demonstrates that MOFs are promising candidate materials not only for low-κ electronics applications but could also be pioneering for terahertz (THz) applications, such as next-generation broadband communications technologies.

Metal-organic framework (MOF) materials have long been considered as promising candidates for industrial gas storage and gas separation due to their significantly large internal surface areas, in the range of 1,000 to 10,000 m$^2$ g$^{-1}$.[1-2] However, large quantities of the materials are required, and this, along with other restrictions such as poor processing abilities, has resulted in the recent shift in scientific focus to investigate the potential of these hybrid materials for next-generation applications, such as drug delivery and sensing.[3-8]

Materials, such as MOFs, featuring "ultra-low-κ" dielectric constants (κ ~ 1.1 to 1.5) and tuneable structural properties and porosity, are ideal candidate materials for the future microelectronics industry. This is due to the dependence of low-κ dielectric response upon porosity, which is highly tuneable in MOF materials.[9]

Of specific interest are devices that can lead to integrated circuits, in which the MOF acts as an interlayer dielectric insulating material.[10] There are, however, some requirements for the engineering of functional devices including new low-κ materials for real-world applications.[11] These include thermal stability at high temperature, predictable mechanical behaviour and long term stability, electrical insulation, and excellent adhesion to other interlayers.

The current literature highlights the promising future of MOFs as low-κ dielectric materials, though little work has been done to tackle any of the above restrictions especially in the terahertz (THz) and infrared (IR) frequency range. Theoretical work in the field exists but has been mainly limited to the semi-empirical Clausium-Mossitti model, and only applied to the static dielectric constant of cubic Zn-based MOFs.[12] There have more recently been some interesting *ab initio* studies on cubic frameworks and the complex dynamic dielectric constant in the near-ultraviolet (UV) region.[13] These theoretical studies have encouraged spectroscopic ellipsometry (SE) to determine the dielectric and optical properties of HKUST-1 thin films.[14] The results were in good agreement with theory, but again the work reported was restricted to the near-UV region.[14] Thin films of ZIF-8 have also been studied in the near-UV region, highlighting zeolitic imidazolate frameworks (ZIFs),[15] as promising candidates as low-κ dielectrics.[16] Other recent studies have investigated the dielectric properties of another Zn-based MOF in the presence of solvent molecules, illustrating the opportunity to tune host-guest interactions[17] and an additional Zn-based MOF that shows high thermal stability.[18]

One major limitation delaying these promising materials for low-κ electronics applications is that experimental methods to study these properties across the IR spectral frequency range are insufficient. This is even more true for the topical THz spectral region, important for emerging broadband communication technologies and applications such as optical sensors.[19] In addition, theoretical methods are only starting to provide an accurate procedure to allow for calculations of the static dielectric constant, as discussed later in this letter.

Fourier Transform IR (FTIR) spectroscopy has been very informative in probing the structural dynamics of MOFs. We recently demonstrated this using synchrotron-based far-IR spectroscopy and inelastic neutron scattering (INS), in conjunction with *ab initio* density

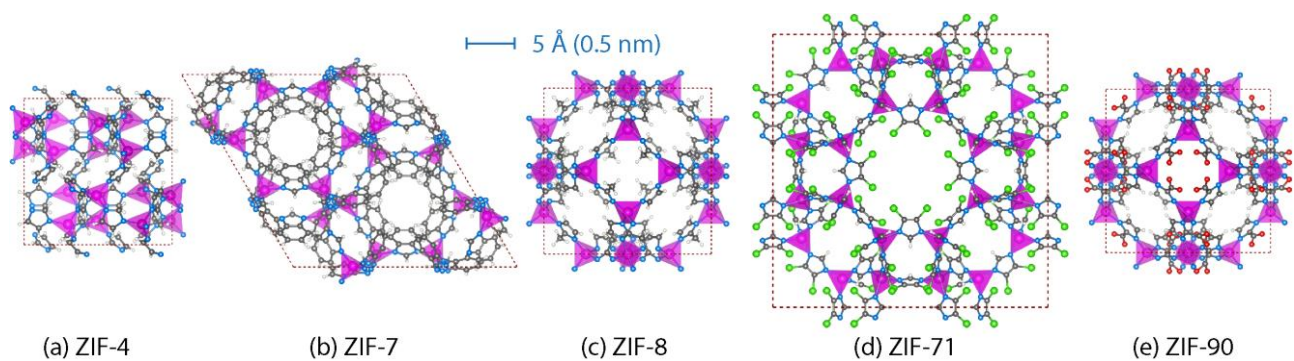

Figure 1. Framework structures of (a) ZIF-4, (b) ZIF-7, (c) ZIF-8, (d) ZIF-71 and (e) ZIF-90, along the crystallographic *c*-axis. The inorganic building units are ZnN$_4$ coordination tetrahedra, highlighted in purple, connected via a series of imidazolate-derived organic linkers. Dashed lines represent one unit cell. Colour scheme adopted: Zn: pink; C: grey; N: blue; O: red; Cl: green; H: white.

functional theory (DFT) to explain the framework specific THz region collective vibrational motions of a variety of MOF structures.[20-22] These low-frequency THz vibrational motions were linked to various physical phenomena, including gate-opening and breathing mechanisms in ZIF-8 and ZIF-7, and shear-driven destabilisation motions for multiple ZIF structures.[23-24] The latter result has been linked to the mechanical induced amorphisation that is observed for ZIF pellets prepared under significant pressure.[20, 25-27]

Synchrotron IR spectroscopy can also be utilized to obtain high spectral quality reflectance spectra across the whole IR spectral range using a single setup. Therefore, avoiding systematic errors. By Kramers-Kronig Transformations (KKT) the real and imaginary parts of the optical properties such as frequency-dependant dielectric constant and refractive index can be easily obtained. In fact, the complex refractive index and dielectric constant are related via KKT as a direct result of the causality principle. The values discussed in this letter were all calculated using KKT implemented by the Bruker OPUS code. The reflectivity of the air/sample interface (*R*) is the square modulus of gamma (Γ) and can be calculated using the Fresnel equation:

$$\Gamma(\nu)e^{i\phi(\nu)} = \frac{n(\nu)-1}{n(\nu)+1}$$

where $\phi(\nu)$ is the phase rotation angle calculated from the frequency (wavenumber). The real and imaginary parts of the frequency-dependant (dynamic) refractive index can then be obtained from:

$$n(\nu) = \frac{1-R(\nu)}{1+R(\nu)-R^{1/2}(\nu)\cos(\phi(\nu))}$$

In this Letter, we report for the first time the dynamic infrared region dielectric constant and refractive index values for five topical polycrystalline ZIF materials (Figure 1). The focus of the Letter is on the dielectric constants, as the static dielectric constant is simply the square of the static refractive index ($\kappa = n^2$). We used synchrotron FTIR specular reflectance spectroscopy (Figure 2) to obtain high signal-to-noise reflectivity data of each material at room temperature and via KKT, we were able to extract the complex dielectric constant (Figure 3). Specular reflectance is the direct reflection from the outermost surface. The reflectance spectra were obtained from solid pellets of each material, using the setup depicted in Figure 2.

The experimental spectra were obtained using a customised setup recently commissioned utilizing a Bruker Vertex 80V Fourier Transform IR (FTIR) Interferometer, to provide *in situ* IR data using a synchrotron radiation source. The advantage is the ability to use specular reflectance which simplifies the KKT calculations and the broadband of the synchrotron radiation to cover the whole IR spectral range. A more detailed description of the experimental setup is available in the Supplementary Information.

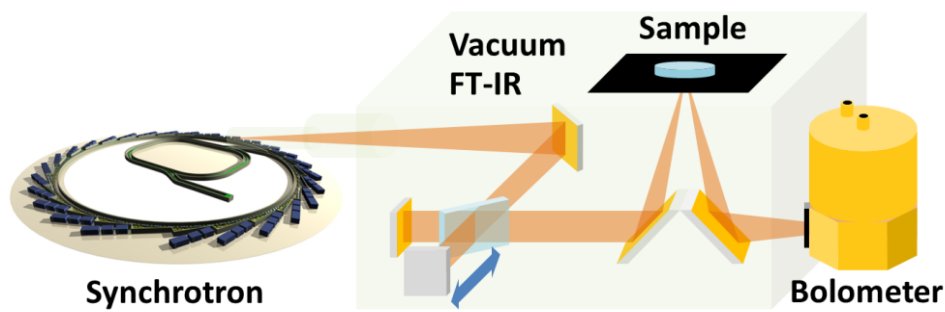

Figure 2. Schematic of the bespoke experimental setup used for the synchrotron reflectivity measurements conducted at the B22 Beamline (MIRIAM) at Diamond Light Source.

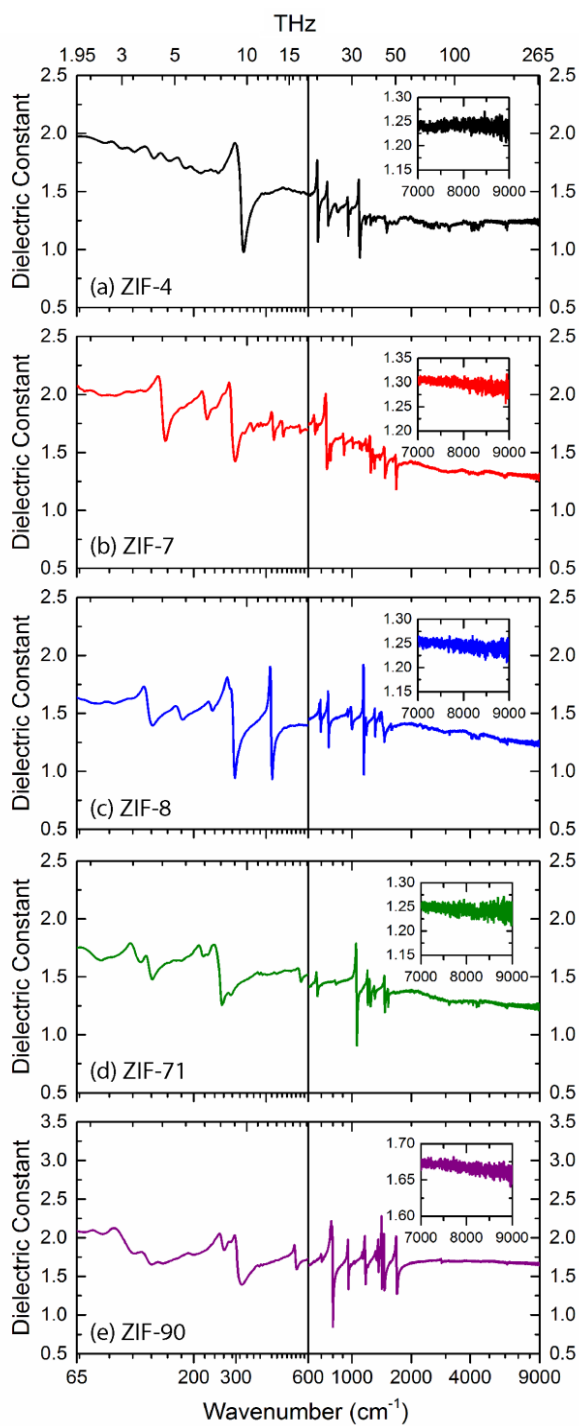

Figure 3. Spectra of the far-IR and mid-IR spectral regions of (a) ZIF-4, (b) ZIF-7, (c) ZIF-8, (d) ZIF-71 and (e) ZIF-90. The spectra show the frequency dependent dielectric data, obtained via Kramers-Kronig transformation. The far-IR and mid-IR regions are both plotted on separate logarithmic axes for additional clarity. The spectral region at the boundary between the mid-IR and near-IR (7000-9000 cm$^{-1}$) is shown as an inset to show the dielectric values more clearly.

KKT is valid and produces reliable results under specific conditions. If the material is absorbing of infrared radiation (as is the case with most materials, including MOFs), then the reflectance spectra will be affected by the absorption bands, and it is, therefore, critical to ensure that the selected spectral cut-offs for the KKTs do not coincide with any absorption features. This is not a major concern towards the near-IR region, as overtone absorptions are very weak. However, care must be taken when selecting the spectral cut-off in the mid-IR and far-IR regions where there are a significant amount of absorption features.[20, 28]

Also, reflection takes place at the surface, requiring the samples to exhibit a low roughness compared to the wavelength (< 100 nm). We therefore ensured that the ZIF pellets were as smooth as was practically possible. This was investigated using an Alicona InfiniteFocus 3D Profilometer which confirmed that the root-mean-square (RMS) roughness of the pellets was ~40 nm and importantly this was not affected by the magnitude of pressure used to press the specific pellet (see Supplementary Information).

We report the first example of experimental frequency dependent dielectric properties of MOF materials in the extended infrared (IR) spectral region (65-9000 cm$^{-1}$). The dynamic dielectric data was obtained for five ZIF structures, namely: ZIF 4, ZIF 7, ZIF 8, ZIF-71, and ZIF-90 which are all Zn-based materials with different imidazolate derived linkers. The crystal structures are plotted in Figure 1. We observe that all dielectric constant values, excluding absorption effects, are within the range of 1.2–2.5. Of particular interest are the low-κ values witnessed closer to the Near-IR spectral region, i.e. beyond 4000 cm$^{-1}$. These values are ~1.2 for all the structures excluding ZIF-90. This is likely due to the structural disorder present in ZIF-90, and for this reason, it will be excluded from the discussions later in this manuscript regarding systematic trends. The near-IR spectral region appears insensitive to framework structure with an overall common asymptotic behaviour.

The spectral data shows that the dielectric constant (excluding absorption bands) reduces with increasing frequency, therefore confirming that the static dielectric constant represents the maximum for each of the ZIF materials. Also, the gradient in the decrease of the dielectric constant with respect to wavenumber is higher in the low-frequency Far-IR region, compared with the Mid- and Near-IR (Figure 4(g)). The data reported in Figures 3 and 5 concerns the fact that the Far-IR dynamic dielectric constants are significantly structure dependent and show a direct link to the porosity and hence the solvent assessable volume of the framework. The direct link to the level of porosity is to be expected from a fundamental point-of-view, as the dielectric constant is a measure of the ability of a substance to store electrical energy in an electric field and as the porosity increases the dielectric constant will tend towards the value of a vacuum (κ = 1) or air (κ ~ 1).

To perform the specular reflectance experiments, the MOF materials were pressed into ~1 mm thick, 13 mm diameter circular pellets using a hydrostatic press. The effects of pelletization were therefore investigated, to ensure that accurate dielectric constant values were being reported and ensure that no pressure induced amorphisation or phase transitions, as often reported for the ZIF family, had occurred.[29]

We performed an extensive investigation into the effects of pressure on the ZIF-8 material. As can be seen from Figure 4, there is a linear trend in the increase of the dielectric constant at each spectral region with increased pelletization pressure. Noteworthy, the pelletization pressure had a greater effect on the THz/Far-IR spectral region with an increased gradient (slope) of 0.71 GPa$^{-1}$, more than double that of the near-IR region (gradient = 0.35 GPa$^{-1}$).

The increase in dielectric constant was confirmed to be primarily an effect of the structure reacting to the pelletization pressure and an additional contribution resulting from the higher densification of the pellet (as mentioned below). This was confirmed via XRD, linking the increase in dielectric constant to the decrease in the Bragg intensities (shown in the Supplementary Information for ZIF-8). We observed the onset of the pressure induced phase transition of ZIF-8 at pressures above 0.300 GPa, indicated by the evolution of a new spectral feature at ~385 cm$^{-1}$ (further highlighted in the Supplementary Information). The spectral feature observed at ~450 cm$^{-1}$ is a numerical artefact resulting from the spectral range used for the KKT. The removal of the artefact is discussed in the Supplementary Information.

Pellets pressed with 0.038-0.150 GPa of pressure (0.5-2.0 tonne over a 13 mm diameter pellet) did not affect the resultant dielectric values significantly, as can be seen from Figure 4(g), and values obtained using these pellets were therefore used to analyse structural trends in the ZIF series (Figure 5).

We observe a direct linear dependency on the THz region dielectric constant with framework solvent assessable volume (SAV) and framework density. However, the experimental values are slightly lower than the theoretical dynamic values of an idealised single crystal, due to the polycrystalline nature of the pellets. The dynamic spectra obtained via DFT is reported for ZIF-8 in Figure 4(a). This minor disparity can be solved by treating the experimental THz dielectric constants on a similar platform to the theoretically calculated static values (effectively zero frequency). Another major advantage of using the theoretical static values is that the dielectric response can be obtained for additional ZIF structures providing for a more accurate trend in the structure-property relationship shown in Figure 5. The comparison between the DFT values (static) and the experimental values (THz) is remarkably good. This, therefore, allows us to predict the THz-frequency dielectric constant (and refractive index) of any ZIF structure based on the structure-property trends highlighted in Figure 5. The porosity and SAV values were calculated using the "VOID" algorithm implemented via the PLATON code.[30]

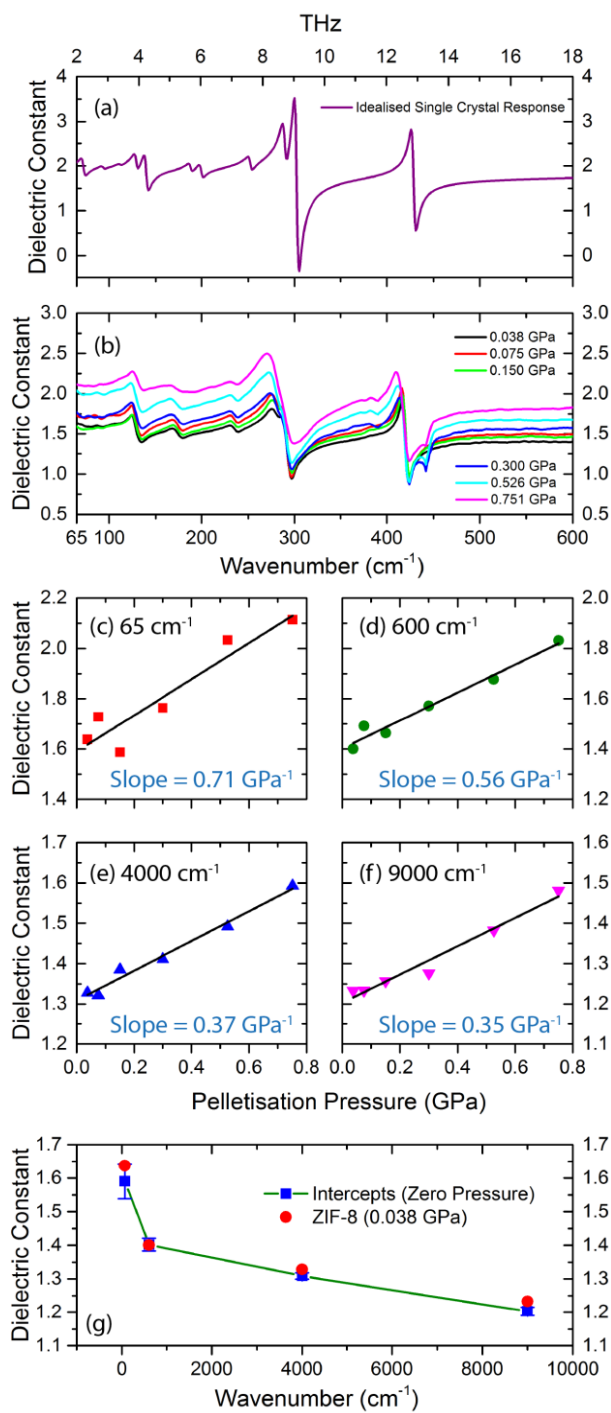

Figure 4. Spectra of the (a) theoretical and (b) experimental far-IR dielectric data of ZIF-8 showing the step-wise effect of pelletization pressure. Plots showing the change in the dielectric constant of ZIF-8, upon increased pelletization pressure, at a range of specific spectral points: (c) 65 cm$^{-1}$; (d) 600 cm$^{-1}$; (e) 4000 cm$^{-1}$; (f) 9000 cm$^{-1}$. The gradient of the trends highlights that the dielectric constant is more reactive to increased palletization in the THz-frequency region (1.95 THz ≈ 65 cm$^{-1}$). (g) The dielectric constant values for ZIF-8 at 65, 600, 4000 and 9000 cm$^{-1}$, extrapolated to assume zero pressure effects.

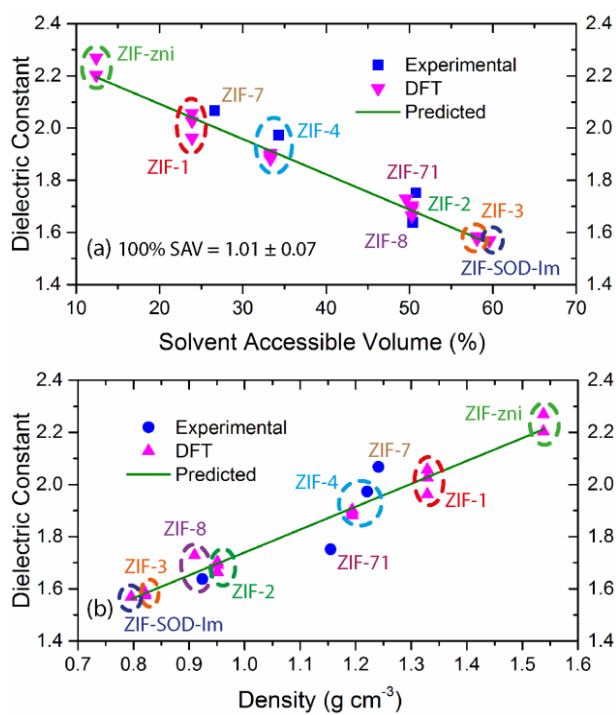

Figure 5. The trends observed in the THz-frequency (1.95 THz ≈ 65 cm$^{-1}$) dielectric constant compared with (a) solvent accessible volume (SAV) and (b) framework density. The experimental SAV and density values are calculated from the experimental CIF files using the PLATON code.[26] The theoretical values are obtained from optimised structures at the B3LYP-D* level of theory. The theoretical dielectric constants are static values (low frequency or constant electric field).

While all the dielectric constant values considered in this Letter are for Zn-based materials, we expect the effect of the specific metal node to be less significant than the steric effects of the specific linkers. Hence, we consider the primary contributing factors to the dielectric constant are the levels of porosity and degree of solvent accessible volume which is related to the framework density (although not linearly). If we extrapolate the trend to a SAV value of 100%, it will result in a dielectric constant of ~1. This is to be expected as a SAV of 100% is essentially air. Therefore, if we consider some of the most porous MOFs, then we can predict dielectric constants of approximately the same response to air. For example the Zn-based MOF with the lowest density obtained from the hypothetical MOF database,[31] would give a dielectric constant of ~1 (density = 0.118 g cm$^{-3}$). In addition, if the choice of metal indeed does not make a major difference then the recently synthesised and lowest-density MOF to be experimentally proven, uranium-based NU-1301 (density = 0.124 g cm$^{-3}$)[32] would also have a dielectric constant of ~1.

This Letter demonstrates for the first time the use of specular reflectance spectroscopy to obtain experimental frequency-dependent (dynamic) dielectric properties and refractive index values of MOFs in the critical infrared (IR) spectral region (65-9000 cm$^{-1}$). Of particular novelty are the THz region dielectric response and the surprisingly good agreement with theoretical static values calculated via DFT. The dynamic dielectric data for the four ZIF structures, namely: ZIF-4, ZIF-7, ZIF-8 and ZIF-71 were analysed, and the THz-frequency dielectric constants were found to be highly structure-dependent and have a linear trend linked to the framework density and its associated SAV. The structure-property trends highlighted in this letter can now be exploited to target future research concerned with dielectric properties of MOFs to particular subgroups and sets the foundation for the promising field of low-κ dielectric porous framework materials and there integration into electrochemical polycrystalline thin film devices.[33-34]

**Acknowledgements**

M.R.R. would like to thank the UK Engineering and Physical Sciences Research Council (EPSRC) for a DTA postgraduate scholarship and also an additional scholarship from the Science and Technology Facilities Council (STFC) CMSD Award 13-05. M.R.R would also like to thank the EPSRC for a Doctoral Prize Fellowship and the Rutherford Appleton Laboratory (RAL) for access to the SCARF cluster and additional computing resources. T.D.B would like to thank the Royal Society for a University Research Fellowship. We are also grateful to the ISIS Support Laboratories, especially Dr Marek Jura and Dr Gavin Stenning at the Materials Characterisation Laboratory (R53) for providing access to X-ray diffraction equipment. The experimental work was performed at large scale facilities through the Diamond Beamtime at B22 MIRIAM (SM10215 and SM14902).

*Email address: jin-chong.tan@eng.ox.ac.uk

**Author contributions:** M.R.R, G.C and J.C.T. designed the research; M.R.R, C.S.K, M.D.F and G.C performed the IR spectroscopy; M.R.R and B.C performed the theoretical calculations; T.D.B, Y.S and I.F synthesised the materials; M.R.R and Z.Z prepared the pellets; M.R.R and K.T performed the XRD work; M.R.R analysed the data; M.R.R wrote the paper with input from other authors.